# OntoForms: User interface structure from a domain ontology

Bruno Szilagyi[1][0009−0004−0522−3876], Edelweis Rohrer[2,3][0000−0002−8257−5291], and Regina Motz[3][0000−0002−1426−562X]

Facultad de Ingeniería, Universidad de la República, Julio Herrera y Reissig 565, 11300 Montevideo, Uruguay
bruno.szilagyi.ibarra@gmail.com
{erohrer,rmotz}@fing.edu.uy

Abstract. This paper presents a software component that generates a user interface structure for populating a domain ontology. The core of this work is an algorithm that takes an ontology and returns a structure describing the user interface. The component also provides functions for populating the ontology and editing existing individuals. Unlike previous approaches, this method can be implemented without any configuration. Additionally, it offers an easy-to-use configuration mechanism that allows irrelevant classes to be hidden and automatically populated. What distinguishes this work is that, instead of exploring the ontology using syntactic methods or queries, our algorithm employs services that implement description logic inference mechanisms. This work illustrates the proposed approach using the well-known wine ontology.

## 1 Introduction

Ontologies are proven modelling artefacts for conceptualizing different domains, as education and health, providing a structured framework to describe them in terms of concepts, instances, and roles (binary relations), as well as constraints about them [8]. Ontologies promote the understanding of the domain, facilitate the formulation of queries on domain data, and have the capability of entailing implicit knowledge from the explicitly declared. Hence, nowadays there are increasingly ontology-based applications that take advantage of the benefis of ontologies, mainly regarding their ability to represent the domain at a conceptual level, hiding the complexity and implementation details of data structures of the domain instances.

From the point of view of the usability quality attribute, an ontology-based application must provide the user with a friendly interface to query and enter data from this conceptual view provided by the ontology. However, as new user needs arise, besides modifying or extending the ontology, the user interface must also evolve. To avoid modifying the application frontend and the logic that populates the ontology every time the ontology changes, there exist approaches that generate data entry forms from ontologies, favouring the maintainability and reusability of ontology-based applications.

The main contribution of the present work is the implementation of a user interface generator component, Ontoforms, which has as a crucial part the algorithm that receives a domain ontology, and optionally a configuration, and returns a structure describing the user interface. Ontoforms also solve the population of the ontology with the values entered by the user. Ontoforms provides domain applications with different endpoints that solve functionalities like obtaining the interface model and populating the ontology once the user enters the data. What distinguishes this component from existing ones is the generation strategy, which allows the generation of the form both with and without configuration, and automatically locates prompts and fields grouped in different sections when necessary, and creates instances that are transparent to the user who enters the data (e.g. instances of classes added to the ontology to represent n-ary relations when n is greater than 2). A key point in this work is the use of reasoning services based on the formal semantics of description logics, which ensures that besides declared axioms, entailments are also obtained and included in the generated form. Moreover, the behaviour of OntoForms is showed by applying it to the well-known wine ontology.

The remainder of this paper is organized as follows. Section 2 provides an overview of related work by contrasting it with the proposed approach. Section 3 describes the architecture of the solution, whereas Section 4 deepens into the details of the core algorithm implemented in Ontoforms. Section 5 illustrates the approach by applying it to a domain ontology and Section 6 presents some conclusions and future work.

## 2 Related work

The work of Gonçalves et al. was conceived to generate UI forms for the health domain even though it can be applied to any domain [3]. The proposed system has three main components: an XML configuration file used for the generation of the web form, a domain ontology, and a data model that describes the form. This approach requires considerable configuration effort from an administrator user since besides the domain ontology it is necessary to provide the XML file and also a data model to configure the form. This also affects the solution maintainability as the configuration file must be reviewed when the domain ontology evolves. Moreover, the generated form allows the user to enter data (ontology instances) but does not enable the update of data already entered. To use the provided implementation, it is necessary to upgrade it because it uses old technology.

The thesis of Liu is also an ontology-based approach that relies on a domain ontology and a called "technology level ontology" which maps the domain ontology to the underlying data schema where the ontology is stored [4]. The interface is generated from the set of attributes that the domain expert selects in the data schema through an intuitive interface. However, the fact that two different ontologies must be maintained can be tedious for the administrator user. This work focuses on drawing the form but not in populating the ontology with the instances entered by the user. This approach also uses old technology. Based on the work of Gonçalves et al., the approach of Vcelak et al. consists of the retrieval and edition of RDF data via web responsive forms generated from an ontology, enabling both the insert and the update of instances already entered [9]. Instead of maintaining a configuration structure, for this purpose, the authors add annotations to the ontology. Even though there is no configuration interface for the administrator user, the configuration of the form does not appear as a tedious task since it consists of just adding annotations, which makes the solution quite maintainable. Even though the authors give a clear description of the approach, the implementation of the solution is not available. Another approach similar to the one of Gonçalves et al. is the proposed by Rutesic et al., for the domain of pizzas [7]. Besides the domain ontology FOODON, an ontology that describes the graphical user interface is a key component of the approach. On the one hand, it is a simple implementation that requires just change the interface ontology when the domain ontology changes, and that generates a user-friendly interface. Besides entering new instances, this interface allows adding new classes or properties to the ontology, which is not supported by Ontoforms. On the other hand, the configuration is too much manual for the administrator user, and the update of instances is not soundly solved. The approach is implemented by using up-to-date technology. Within the manufacturing domain, the quality of manufacturing service capability (MSC) information can be measured by its accuracy and semantic interoperability. For this, the use of ontologies that describe MSC information contributes to a correct understanding among all stakeholders, avoiding ambiguity. To help manufacturing domain experts (without expertise on ontologies) to provide MSC information using the ontology vocabulary, Peng at al. propose the Ontology-based eXtensible Dynamic Form (OXDF) user interface architecture [5]. Some relevant functionalities of this approach are the generation of forms from ontologies, a search engine to find ontology entities and a mechanism for the user to insert and update instances, as well as to de ne new classes or properties, which is not enabled by Ontoforms. This approach does not allow configuring the form to select the classes or properties that are presented or hidden. As the generated interface presents the whole ontology, the solution is easy to evolve. The implementation of the approach is not available. The work of Rocchetti et al. presents a general ontology-based approach that is applied to the art domain. Besides configuring some aspects such as the set of key ontology concepts from which the form is generated, a graph representing the ontology structure is constructed as an intermediate step to generate the form, which allows the insert of instances but not the update of existing ones [6]. An intuitive configuration interface is provided to the administrator user. The form generation takes this configuration as input, so the solution easy to evolve. The implementation of the approach requires to be upgraded. Even though Ontoforms takes ideas of the works described above, it presents some advantages. Ontoforms provides an easy-to-use configuration interface for the administrator user, where the user selects the configuration options by just selecting from the hierarchy of ontology classes and properties. As a plus, Ontoforms provides a function to generate a basic ontology form without any configuration. Only the work of Peng at al. does not require previous configuration, but it does not provide any configuration mechanism. Ontoforms enables the insert and update of instances, which is also provided by some of the selected related work, whereas others only enable the insert of instances. Since application requirements usually change over time, the ontology frequently changes as well. In this case, the implementation of Ontoforms does not need to be mod i ed. Instead, the administrator user will have to revise the form configuration through the interface provided. Ontoforms differs from existing approaches in the use of reasoning services based on the formal semantics of description logics, to ensure that both declared and entailed axioms are obtained to draw the user interface. Moreover, the implementation of Ontoforms is modular, and uses up-to-date technologies.

Table 1 compares the different approaches, including Ontoforms, regarding a set of selected quality attributes. Some quality attributes in Table 1 are relevant for users, both administrator users (in charge of ontology and form configuration tasks) and end users (users that will use the generated form to enter and query data), whereas other quality attributes address the internal quality of the solution. For administrator users we consider the configuration e ort, to evaluate how complex the configuration task is. For the end user, the usability for data editing is selected to evaluate how friendly the behavior of the user interface is, which considers if the form only enables the insert of instances or also allows the update of instances, and if the solution implements the automatic generation of ontology instances without meaning for the user (e.g. instances of intermediate classes to represent n-ary relations). Another quality attribute relevant for the end user, in particular for expert users, is the capability of adding classes and properties, which enables users without expertise in ontology engineering to change the ontology structure. Regarding the internal quality of the solution, the maintainability attribute is considered, defined as "the degree of e ectiveness and e ciency with which a product or system can be modified to improve it, correct it or adapt it to changes in environment, and in requirements", which includes the modularity and reusability of the software [2]. Moreover, it is relevant the technology upgrading, to evaluate how up-to-date the implementation technology is. Finally, as described above, we consider the use of reasoning services to obtain the ontology axioms to generate the form; besides affecting the internal quality of the solution, this can influence what the user interface shows. Table 1 shows the comparison of approaches for the selected quality attributes through the set of values good, acceptable and bad, where the value good is assigned if the evaluation of a quality attribute is positive for the user or the internal quality of the solution (as appropriate), the value bad is assigned if it is clearly negative, and the value acceptable when it is in the middle.

| Approach | Con g. e ort | | Adding classes and properties | | Mai T | upgrading |
|---|---|---|---|---|---|---|
| Gonçalves et al. | good | bad | bad | bad | bad | bad |
| Liu | acceptable | bad | bad | bad | acceptable | bad |
| Vcelak et al. | good | bad | bad | bad | acceptable | not avail. |
| Rutesic et al. | bad | bad | good | bad | acceptable | good |
| Peng at al. | not appl. | good | good | bad | good | not avail. |
| Rocchetti et al. | good | bad | bad | bad | good | bad |
| Ontoforms | good | good | good | good | good | good |

Table 1. Comparison of approaches

## 3 Overview of the OntoForms architecture

This section sketches the general solution architecture that takes advantage of the OntoForms software component. OntoForms improves the maintainability of an ontology-based application since when the application requirements and (as a consequence) the ontology evolve, the user interface can be regenerated by OntoForms. Besides the domain user, who interacts with the application to play his/her role in the domain, the solution also considers the administrator user who is in charge of uploading ontologies, generating corresponding forms, and con guring them if necessary.

Figure 1 illustrates the proposed architecture.

The left side of the figure shows the OntoForms System which provides serices to the administrator user. The frontend application Ontoforms Admin provides the user an interface to interact with the Ontoforms System, making use of different endpoints provided by the Ontoforms Core API, which is the main component of the architecture. The Ontoforms Core API solves different functionalities, like obtaining the ontology from the Triplestore and generating the structure that describes the ontology data entry form. Moreover, it provides endpoints to upload new ontologíes and save form configurations, among others.

The right side of Figure 1 shows the Domain System, that is accessed by the end user. Even though the implementation of a domain application can be solved following different architecture styles or patterns, for the sake of illustrating the idea of a complete solution, we again propose a layers architecture style, with a frontend, named Domain Assistant App, and a backend, the Domain Assistant Api. As the figure shows, the key here is that the backend will also make use of the endpoints provided by the Ontoforms Core API to solve many requirements. Ontoforms Core API provides the backend the structure for the application frontend to draw the user interface. Once the end user enters the data, the backend uses the Ontoforms Core API endpoint that populates the ontology with entered instances and also with intermediate instances that are not added by the user, but that are part of the ontology structure.

The main contribution of the present work is the algorithm that receives the ontology and returns the structure that describes the user interface to populatethe ontology. This algorithm is implemented in the Ontoforms Core API. It is described in detail in the next section.

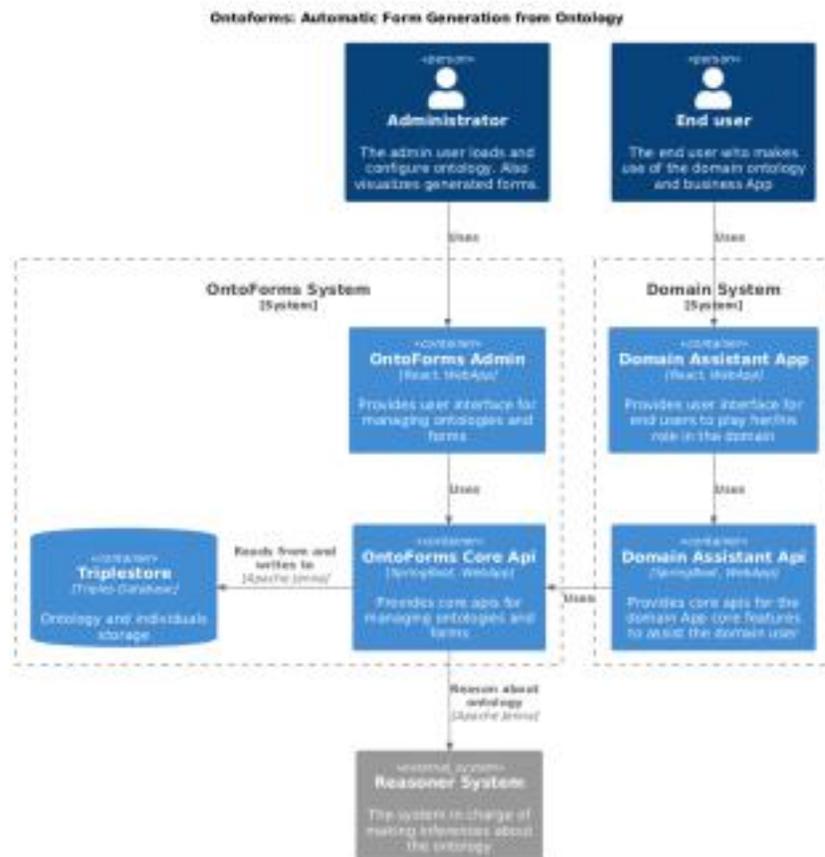

Fig. 1. Architecture of the solution

## 4 OntoForms Core API

The OntoForms Core API is the component that works on top of an ontology to generate the form structure for entering individuals in each class. It operates within a given main class to generate the form structure. This form structure will correspond to the ontology's subgraph accessible from that class through its declared properties and the properties inferred by the reasoner. This process involves filling in all data properties and object properties associated with the individual that populates the main class. Thus, ontology processing is performed in the following steps.

For each data property relation where the main class is within the domain of the relation, OntoForms Core API creates a field in the form structure to enter or select the corresponding data.

For each object property relation where the main class is within the domain of the relation, a section is created in the form structure to enter the new individual related to the individual in the main class by that relationship. This new individual is entered recursively by setting the class within the range of the object property relation as the main class. This processing is depth-first, similar to traversing a spanning tree in the ontology graph. If there are subclasses with this class as a parent, the application allows selecting the most specific subclass as the main class before continuing the recursion. The object property relation is ignored when the main class is part of a complex expression in the object property domain that uses the conjunction operator with other classes. The reason for this is that when the domain is defined as a conjunction of classes, there can be individuals of each of the classes that are not members of the restricted domain. If individuals were to be specifically included in the domain of this relationship, a class defined as equivalent to the expression of that domain should be placed as the main class.

If the individual to which the main class individual is to be linked via the object property already exists, the application will open a selection box on the form. If the object property is functional, then this box will only allow a single selection. Otherwise, the box will allow more than one individual to be selected.

The application allows to ignore, through user declarations, the generation of entries in the form structure for the creation of individuals or the input of data that can be derived.

OntoForms Core API takes the names of the form sections and fields from existing annotations and names in the ontology. However, these names can be customized.

**Implementation Issues**

One of the challenges of ontology processing is retrieving the set of object property relationships defined and inferred for a given class. Retrieving these using SPARQL involves writing queries with complex logic. As described in [1], one consequence of the RDF design is that, because of the open-world paradigm, it is not possible to answer the question "What properties can be applied to the resources of class X? Strictly speaking, the RDF answer is "any property". But to answer the question of which properties have a particular class in their domain, one must also take into account properties that do not have their domain defined, and properties that have the closure of the parent classes in their domain on the hierarchy of classes of the given class.

In this work, to obtain the properties of a given class, we use a mechanism built into Jenna that supports a frame-like view of resources using the reasoner's capabilities. It defines properties of a class as only those properties that don't add any new information when applied to resources already known to be of that class. These properties are the relationships that need to be filled in on a new individual form of the main class. For this purpose, we used Apache Jena SDK to load an ontology model with one of the configurable reasoners (we use the

OWL_DL_MEM_TRANS_INF reasoner configuration), and then retrieve from the target ontClass the declaredProperties filtered with filterKeep of data and object properties.

## 5 World Wine ontology example

In this section, we describe the application of the OntoForms Core API to the Wine ontology to produce the user interface form structure. We show the Onto Forms Core API administrator web page and preview the generated form structure suitable for adding a new individual to a selected main class. Figure 2 shows some classes and properties of the Wine ontology.

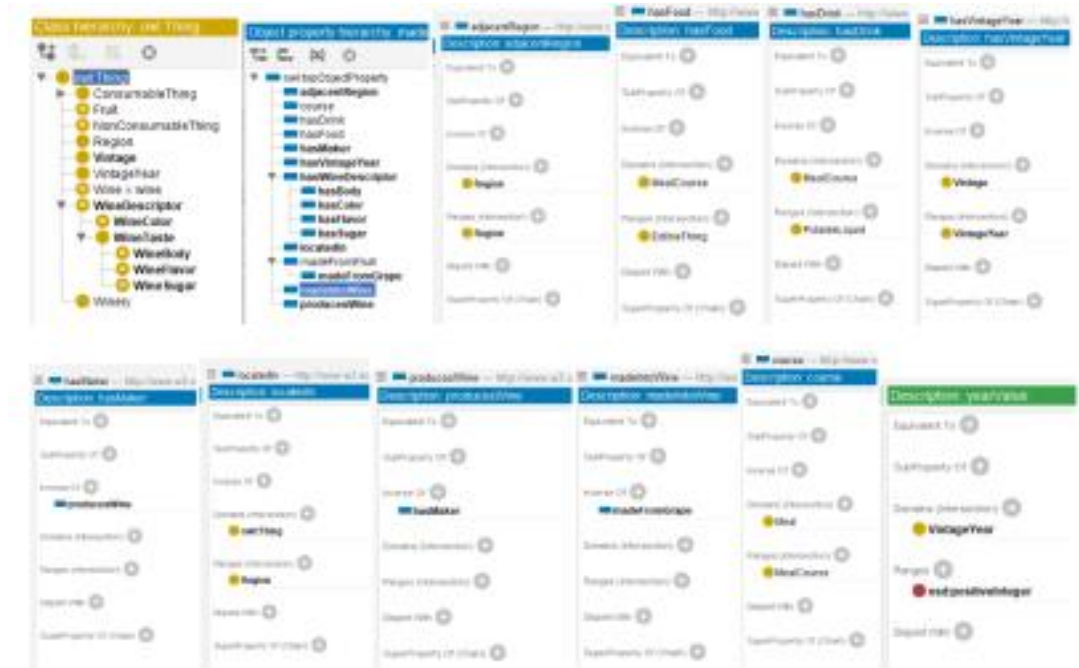

Fig. 2. Overview of the Wine Ontology

To visualize a form to populate the ontology, the first step is to upload the ontology into OntoForms system via the OntoForms administrator webpage (accesible via http://ontoforms-ui.web.elasticloud.uy/), as Figure 3 shows.

Once the ontology is uploaded it can be explored by pressing the Onto detail section. This can be helpful to recognize all of its classes and the hierarchy of them alongside the properties and individuals for every one selected (see Figure 4).

After that, the administrator user can select an ontology and a main class from the class hierarchy tree to display the default form for the selected main class, illustrated in Figure 5. Once the default form is visualized, it can be configured for example by removing properties that do not want to be instantiated when creating a new individual for the class.

For this, the configuration menu is available, in this example the object properties hasMaker, MadeIntoWine, and ProducesWine are removed from the form (see Figure 6). Next, the user can decide that it is needed to create a new indiidual instead of choosing from the individuals available for the property course, with range "MealCourse". So instead of rendering a combobox for the "course" property, a new section of the form is rendered for the properties of the class MealCourse, as Figure 9 shows.

Fig. 3. Step 1: Select and upload the Wine Ontology

Fig. 4. Step 2: Explore uploaded Ontology

Fig. 5. Step 3: Previsualize default form for Meal class

Fig. 6. Step 4: Configure property hiding

Fig. 7. Step 4: Result form with hidden properties hasMaker, MadeIntoWine, and ProducesWine

Fig. 8. Step 5: Con gure class no instance choose.

Fig. 9. Step 5: The nal form with section of MealCourse is con gured.

In this case, the algorithm generates a new instance of MealCourse, and associates it to instances selected by the user for the properties with domain MealCourse.

## 6 Conclusions and future work

This work presents the software component OntoForms, which takes a domain ontology as input and returns a structure describing a user interface. Ontology based applications bene t from OntoForms, as their frontend components can use endpoints provided by OntoForms to generate the user interface. This contributes to the maintainability of ontology-based applications.

The core of OntoForms is an algorithm that, given an ontology and a class within it, generates a structure representing a form to insert or update an individual of this class. It also instantiates the object and data properties associated with this individual. The key distinguishing feature of OntoForms is its use of reasoning services based on the formal semantics of description logics, ensuring that both declared and entailed axioms are used to build the user interface.

OntoForms can be used by simply providing the ontology and the class to be instantiated. It also offers a configuration interface that allows the administrator to select which ontology classes and properties should be included or hidden in the form, according to domain application requirements.

Currently, an ontology-based application in the health domain is being implemented to leverage the benefits of OntoForms. The next step is for health experts to validate the usability of domain applications that use OntoForms.